\documentclass[pra,twocolumn,showpacs,amsmath,amssymb]{revtex4-1}
\usepackage{bm}
\usepackage[pdftex]{graphicx}
\usepackage{color}

\begin{document}
\title{Annihilation and recurrence of vortex--antivortex pairs in two-component Bose--Einstein condensates}
\author{Junsik Han$^1$}
\author{Makoto Tsubota$^{1,2,3}$}
\affiliation{$^1$Department of Physics, Osaka City University, 3-3-138 Sugimoto, Sumiyoshi-ku, Osaka 558-8585, Japan}
\affiliation{$^2$The Advanced Research Institute for Natural Science and Technology, Osaka City University, 3-3-138 Sugimoto, Sumiyoshi-ku, Osaka 558-8585, Japan}
\affiliation{$^3$Nambu Yoichiro Institute of Theoretical and Experimental Physics (NITEP), Osaka City University, 3-3-138 Sugimoto, Sumiyoshi-ku, Osaka 558-8585, Japan}
\date{\today}


\begin{abstract}

The annihilation of vortex--antivortex pairs is a key event in two-dimensional Bose--Einstein condensates (BECs).
It is known that dissipation or a catalyst vortex is required for the annihilation of the pairs in one-component BECs.
We numerically confirmed in two-component BECs that the pairs can be annihilated even without any dissipation or catalyst vortices when the intercomponent interaction is strong.
In addition, the pair is recreated alternately in two components under certain conditions, which we call the recurrence.

\end{abstract}

\maketitle


\section{Introduction}\label{sec:Introduction}

Quantum turbulence (QT) has been studied in superfluid helium for more than half a century and in atomic Bose--Einstein condensates (BECs) during the past decades \cite{Tsubota13, Barenghi14, Madeira20}.
QT consists of a large number of quantized vortices, which are well-defined topological defects with quantized circulation.
The core size of the quantized vortex is characterized by the coherence length $\xi$, and the diluteness of BEC gas makes $\xi$ on the order of $\mu \rm{m}$; consequently, optical techniques make the core visible.
In addition, the sign of the circulation of vortices is confirmed by the Bragg scattering method \cite{Seo17}.
Thus, QT is studied by focusing on the dynamics of the vortices in BECs both theoretically and experimentally.

In BEC systems, two-dimensional (2D) QT has been studied.
The 2D QT is realized by using a tightly confining trap potential along one axis and restricting the motion of the vortices in the 2D space.
In 2D QT, the formation of clusters of like-sign vortices has been studied extensively \cite{White12, Neely13, Billam14, Simula14, Reeves14, Yu16, Groszek16, Salman16, Gauthier19, Johnstone19}.
It is thought that this phenomenon is related to the inverse energy cascade \cite{Kraichnan67, Reeves13}, wherein the energy is transferred from a small to a large spatial scale and the vortices create a large structure.

In 2D systems, the dynamics of a vortex--antivortex pair is one of the simplest and most important elements of the dynamics of the vortices.
If the system has no dissipation, a pair is stable and moves with a self-induced constant velocity.
However, a vortex pair can be annihilated through three mechanisms. 
First is through dissipation. 
When the fluid is incompressible, a dissipative mechanism can annihilate a vortex pair \cite{Kim16}. 
Even when the fluid is compressible, a vortex pair maintains the motion without dissipation.
If some dissipation is included, a vortex pair gradually decreases the distance between two vortices and eventually gets annihilated when emitting the density waves. 
Second is through a many-body process. 
When catalyst vortices interact with a pair, they can annihilate the pair. 
This process has been studied in the decay of QT \cite{Nazarenko07, Kwon14, Stagg15, Cidrim16, Groszek16,  Karl17, Baggaley18, Groszek20}.
Four-body (two catalyst vortices) \cite{Cidrim16, Groszek16,  Baggaley18, Groszek20}, three-body (one catalyst vortex) \cite{Nazarenko07, Groszek16, Baggaley18, Groszek20}, two-body (no catalyst vortex) \cite{Kwon14, Stagg15, Cidrim16, Groszek16, Baggaley18, Groszek20}, and one-body (drift out from the trap) \cite{Kwon14, Stagg15, Cidrim16, Groszek16, Groszek20} processes have been studied with and without dissipation. 
Third is through effective dissipation due to phonons \cite{Shukla14}, however, this process can be negligible in our case \cite{one_component}.

In this paper, we present another novel scenario of vortex pair annihilation in two-component BECs. 
Multicomponent BECs have also been studied in two dimensions \cite{Kasamatsu03, Kasamatsu05, Eto11, Nakamura12, Kasamatsu16, Karl13_2} and three dimensions \cite{Takeuchi10, Ishino11}.
In two-component BECs without dissipation, the total energy of the system is conserved; however, they exchange energies through the intercomponent interaction.
Then, we found that a vortex pair can be annihilated even without dissipation or catalyst vortices when the repulsive intercomponent interaction is sufficiently strong. 
Here, the repulsive intercomponent interaction causes the exchange between the energy of each component and the intercomponent-interaction energy.
This energy exchange results in drastic dynamics, namely, the recurrence.
After a vortex pair is annihilated in one component, another vortex pair appears in another component. 
Thereafter, the pair gets annihilated and a vortex pair appears again in the original component. 
The two components repeat the pair annihilation and creation by exchanging the kinetic energy between them.
Whether this recurrence occurs or whether the pair just gets annihilated depends on the system size.
In coherently coupled BECs, the periodic vorticity transfer between the two components is also confirmed \cite{Calderaro17}.
However, this transfer comes from Rabi coupling depending on the phase difference between the two components, which is different from the recurrence in this paper.

This recurrence accompanies the creation of rarefaction pulses, which are dark blobs without vorticity \cite{Jones82, Berloff02, Feijoo17}.
Jones and Roberts studied the 2D Gross--Pitaevski equation with axisymmetric disturbances, and found that the dispersion relation shows a continuous sequence of a vortex pair and a rarefaction pulse \cite{Jones82}.
This study assumed a one-component uniform condensate, although the scenario is also available in such systems as trapped or two-component BECs.
In general, a vortex can be nucleated from a low-density region, e.g., owing to a repulsive potential \cite{Sasaki10, Fujimoto11} or to the outskirt of the trapped condensate \cite{Tsubota02}.
Even in a uniform system, a rarefaction pulse triggers a low-density region, from which a vortex pair appears in a one-component system under some strong disturbance \cite{Berloff02, Feijoo17}.
In two-component BECs, the intercomponent interaction enables the continuous transformation between a vortex pair and a rarefaction pulse; a similar phenomenon is observed numerically in three-dimensional two-component BECs \cite{Takeuchi10}.
Thus, a rarefaction pulse is also created alternately in both components whenever a vortex pair is created through the recurrence.

The rest of the paper is organized as follows. 
In section \ref{sec:Model}, we introduce the numerical model and explain the assumption of the system. 
In section \ref{sec:Results}, we show the two results of our simulation. 
The pair annihilation without dissipation is presented in section \ref{sec:annihilation}. 
In section \ref{sec:recursion}, we describe the creation of the rarefaction pulse and the recurrence of a vortex pair. 
We discuss the distribution of the kinetic energy when a rarefaction pulse or a vortex pair is created in section \ref{sec:distribution_kinetic_energy}. 
Finally, section \ref{sec:Conclusions} provides the conclusions.


\section{Model} \label{sec:Model}

In this paper, we focus on 2D two-component BECs.
The Gross--Pitaevskii(GP) equation can quantitatively describe many experimental results \cite{PethickSmith08, Tsatsos16, Tsubota17}.
The GP equations in two-component uniform BECs are written as
\begin{equation}
\begin{array}{l}
\mathrm{i}\hbar\frac{\partial}{\partial t}\psi_{j}({\bm r},t) \\
 \\
 = \left[-\frac{\hbar^{2}}{2m_{j}}\nabla^{2} + \displaystyle\sum_{j'=1,2} g_{jj'}|\psi_{j'}({\bm r},t)|^{2} - \mu_{j} \right] \psi_{j}({\bm r},t), \\
 \\
   \hspace{1pc}(j = 1, 2)\label{eq:GPEs}
\end{array}
\end{equation} 
In these equations, $\psi_{j} = \sqrt{n_{j}({\bm r},t)}\mathrm{e}^{\mathrm{i} \theta_{j}({\bm r},t)}$ is the macroscopic wave function of the $j$-th component, where $n_{j}({\bm r},t)$ is the density of the condensate and $\theta_{j}({\bm r},t)$ is its phase.
Here, $m_{j}$ is the mass of the atom, $g_{jj'}$ is the intensity of atom--atom interaction, $\mu_{j}$ is the chemical potential, $\hbar =h/(2\pi)$, and $h$ is Planck's constant.
For simplicity, we choose $m_{1} = m_{2} = m$ and $g_{11} = g_{22} = g$.
Throughout the paper, we use Eq. (1) without any dissipation term. 

In this paper, we consider the case in which component 1 contains a vortex pair and component 2 has no pair in the initial state.
We imprint vortices in component 1 by multiplying the wave function by a phase factor $\displaystyle \Pi_{i}\exp(\mathrm{i} \phi_{i})$, with $\phi_{i}(x,y) = s_{i}\arctan[(y-y_{i})/(x-x_{i})]$.
Here, the coordinates $(x_{i},y_{i})$ and $s_{i}$ refer to the position and the sign of the circulation of the $i$-th vortex ($i =1, 2$), respectively.
After the imprinting, the wave function evolves in imaginary time.
When the vortex cores are formed and the distance between two vortices becomes $5.75\xi$, we stop the imaginary time evolution and treat this state as the initial state of the real time evolution \cite{Ruostekoski05, Groszek16}.
Then, stopping the imaginary time evolution leaves the residual energy due to phonons, however, this hardly affects the dynamics studied in this paper.

In this calculation, Eq. (\ref{eq:GPEs}) is made dimensionless by the length scale $\xi = \hbar / \sqrt{mgn}$, with the time scale $\tau = \hbar / gn$, where $n$ is the density of the uniform condensate.
The dimensionless wave function is defined as $\tilde{\psi}_{j}({\bm r},t) = \displaystyle \frac{1}{\sqrt{n}} \psi_{j}({\bm r},t)$, and the unit of energy is $gn$.
We perform these numerical calculations under the periodic boundary condition with three different periods of $L = 64$, $32 $, and $16$.
Here, the space is divided by an $N^{2}$ uniform Cartesian mesh, and we consider the resolution of the space 
$dx = dy = L/N = 0.25$ for $L = 64, 32$ and $dx = dy = L/N = 0.0625$ for $L = 16$.
We solve the dimensionless GP equation using the Fourier spectrum method and the fourth-order Runge--Kutta method.
In the remaining paper, we use the dimensionless quantities.

It is important to divide the total energy of the system into components in the study of quantum hydrodynamics \cite{Nore97, Kobayashi05}.
In the present system, the energy $E_{j}(t)$ of the $j$-th component is composed of the kinetic energy $E_{{\rm kin}, j}(t)$, quantum energy $E_{{\rm q}, j}(t)$, and intracomponent-interaction energy $E_{{\rm int}, j}(t)$,
\begin{equation}
 E_{j}(t) = E_{{\rm kin}, j}(t) +  E_{{\rm q}, j}(t) + E_{{\rm int}, j}(t), \label{eq:2C_Etot}
\end{equation} 
These energies are defined as
\begin{equation}
 E_{{\rm kin},j}(t) = \frac{1}{2}\int \left[ \sqrt{n_{j}({\bf r},t)}\nabla \theta_{j}({\bf r},t) \right]^{2} d{\bf r}, \label{eq:2C_Ekin}
\end{equation} 
\begin{equation}
 E_{{\rm q},j}(t) = \frac{1}{2}\int [\nabla n_{j}({\bf r},t)]^{2} d{\bf r}, \label{eq:2C_Eq}
\end{equation} 
\begin{equation}
 E_{{\rm int}, j}(t) = \frac{g}{2}\int n_{j}({\bf r},t)^{2} d{\bf r}, \label{eq:2C_Eint}
\end{equation} 
The kinetic energy is divided into the incompressible kinetic energy $E_{{\rm kin}, j}^{\rm i}(t)$,  owing to the quantized vortices, and the compressible kinetic energy $E_{{\rm kin}, j}^{\rm c}(t)$, owing to  the density waves, defined by
\begin{equation}
 E_{{\rm kin},j}^{\rm i}(t) = \frac{1}{2}\int \left[ \left\{\sqrt{n_{j}({\bf r},t)}\nabla \theta_{j}({\bf r},t)\right\}^{\rm i} \right]^{2} d{\bf r}, \label{eq:2C_Ekin_i}
\end{equation} 
\begin{equation}
 E_{{\rm kin},j}^{\rm c}(t) = \frac{1}{2}\int \left[\left\{\sqrt{n_{j}({\bf r},t)}\nabla \theta_{j}({\bf r},t)\right\}^{\rm c} \right]^{2} d{\bf r}. \label{eq:2C_Ekin_c}
 \end{equation} 
Here, $\{\cdots\}^{\rm i}$ and $\{\cdots\}^{\rm c}$ denote the incompressible and compressible components, respectively; $\nabla \cdot \{\cdots\}^{\rm i} = 0$ and $\nabla \times \{\cdots\}^{\rm c} = 0$.
In the two-component system, the intercomponent-interaction energy $E_{{\rm int}, 12}(t)$ is given by
\begin{equation} 
E_{{\rm int}, 12}(t) = g_{12}\int n_{1}({\bf r},t)n_{2}({\bf r},t) d{\bf r}, \label{eq:2C_Eint12}
\end{equation} 
and the total energy of the system is $E_{1}(t) + E_{2}(t) + E_{{\rm int}, 12}(t)$.


\section{Results} \label{sec:Results}

\subsection{Vortex pair annihilation in two-component BECs}\label{sec:annihilation}

We consider the periods of the systems $L = 16$, $32$, and $64$, and call these systems the small, medium, and large system, respectively.

In the one-component system without dissipation, we confirm that the pair moves with constant velocity and never gets annihilated as long as the cores of two vortices do not overlap initially. 
Then, the incompressible and compressible kinetic energy and quantum energy are almost constant with fluctuation.

In the two-component system with $g_{12} \lesssim 0.8g$, we confirm that a vortex pair also moves with constant velocity and never gets annihilated like in the one-component system.
In this paper, we show the results with $g_{12} = 0.9g$ to describe the annihilation of the vortex pair and the recurrence.


\begin{figure*}[t]
\begin{center}
\includegraphics[width=40pc, height=14.07pc]{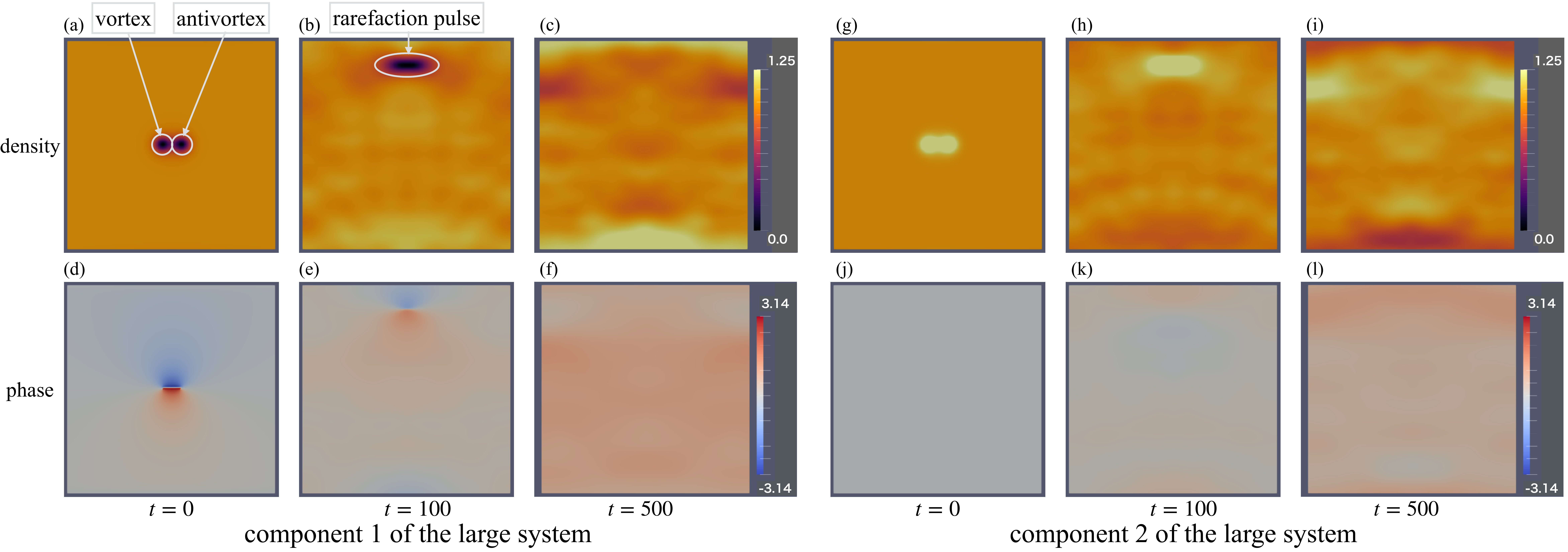}
\caption{\label{fig:2C_5.75_64_n_p} 
Dynamics when a vortex pair just gets annihilated in component 1 in the large system of $L = 64$ with $g_{12} = 0.9g$.
 (a)--(f) describe component 1, and (g)--(l) describe component 2.
(a), (d),  (g), and (j) are the density and phase at $t = 0$; (b), (e), (h), and (k) are those at $t = 100$; and (c), (f), (i), and (l) are those at $t = 500$, respectively.
The color bar of the density is normalized by the uniform density, and the range is $0$ -- $1.25$.
The white circles, ellipse, and arrows denote the vortex, antivortex, and rarefaction pulse, respectively.}
\end{center}
\end{figure*}



\begin{center}
\begin{figure}[h]
\includegraphics[width=20pc, height=16.67pc]{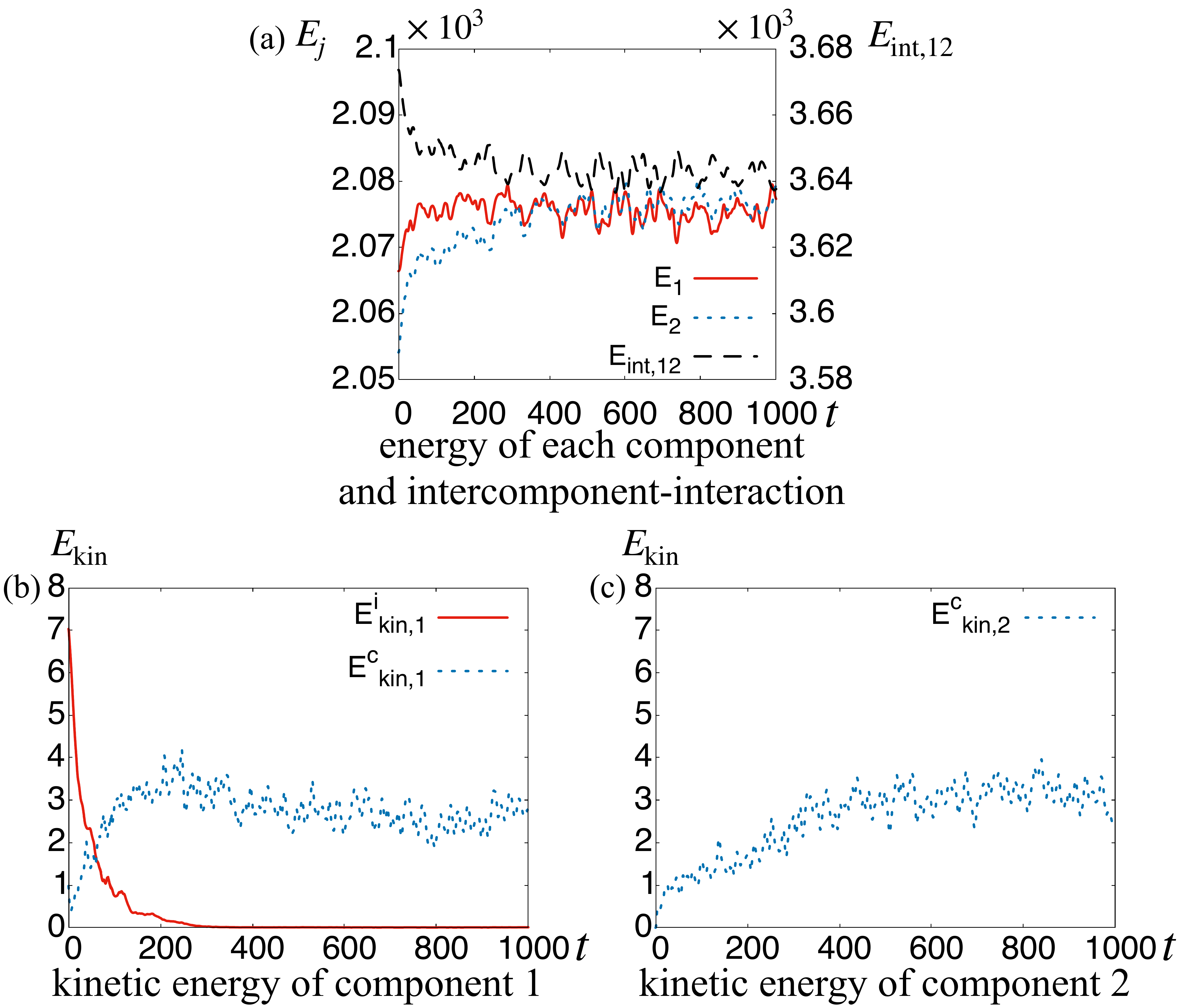}
\caption{\label{fig:2C_5.75_64_d_E} 
Time development of (a) $E_{j}(t)$ and $E_{{\rm int},12}(t)$, (b) $E_{{\rm kin}, 1}^{\rm i}(t)$ and $E_{{\rm kin}, 1}^{\rm c}(t)$, and (c) $E_{{\rm kin}, 2}^{\rm c}(t)$ in the large system with $g_{12} = 0.9g$.
In (a), the red solid, blue dotted, and black dashed lines show $E_{1}(t)$, $E_{2}(t)$, and $E_{{\rm int},12}(t)$, respectively.
In (b) and (c), the red solid and blue dotted lines denote $E_{{\rm kin}, 1}^{\rm i}(t)$ and $E_{{\rm kin}, j}^{\rm c}(t)$, respectively.}
\end{figure}
\end{center}


Figure \ref{fig:2C_5.75_64_n_p} illustrates the density and phase at $t = 0$, $t=100$, and $t =500$ of the large system.
The cores of the vortices of component 1 are filled by component 2 (Fig. \ref{fig:2C_5.75_64_n_p}(a) and (g)) because the intercomponent interaction is repulsive \cite{Eto11}.
In the initial state, only component 1 has a vortex pair where the distance between the vortex and the antivortex is equal to $5.75$.
Then, density waves are not excited in any component, and thus, $E_{{\rm kin}, 1}^{\rm i}(t)$ is dominant in the kinetic energy of the whole system (Fig. \ref{fig:2C_5.75_64_d_E}(b) and (c)).
Here, Fig. \ref{fig:2C_5.75_64_d_E}(c) only shows $E_{{\rm kin}, 2}^{\rm c}(t)$ because $E_{{\rm kin}, 2}^{\rm i}(t) \simeq 0$.
This pair moves upward because of the self-induced velocity.
As the pair moves, density waves are emitted and the pair gets annihilated at $t \simeq 100$, and then a rarefaction pulse appears (Fig. \ref{fig:2C_5.75_64_n_p}(b) and (e)).
Thereafter, the rarefaction pulse decays and the density waves are emitted.
In these processes, $E_{{\rm kin}, 1}^{\rm i}(t)$ is transferred not only to $E_{{\rm kin}, 1}^{\rm c}(t)$ but also to $E_{{\rm kin}, 2}^{\rm c}(t)$ until $t \simeq 300$ (Fig. \ref{fig:2C_5.75_64_d_E}(b) and (c)) because of the intercomponent interaction.
Finally, the density waves spread in both components (Fig. \ref{fig:2C_5.75_64_n_p}(c) and (i)). 
Then, the system becomes statistically steady and vortex pairs never appear.
Through the annihilation of a vortex pair and the spread of the density waves, the phase of component 2 is almost uniform (Fig. \ref{fig:2C_5.75_64_n_p}(j)--(l)).

As the vortex pair gets annihilated, the energies $E_{j}$ of both components increase and the intercomponent-interaction energy $E_{{\rm int}, 12}$ decreases, while the sum of these total energies is conserved (Fig. \ref{fig:2C_5.75_64_d_E}(a)).
These increases and decreases are due to the spread of the density waves emitted by the motion of the pair.

In the one-component system, the pair dose not get annihilated without dissipation or a catalyst vortex.
However, in the two-component system, the pair can get annihilated even without dissipation or a catalyst vortex, which is caused by the energy exchange through the intercomponent interaction.
Here, we describe the details of the exchange of energies.
In the initial state, the densities of both components are uniform except for the cores of the vortices, which are represented by $n_{{\rm uni},j}$.
The spread of the density waves changes the densities as $n_{{\rm uni},j} + \delta n_{j}({\bf r}, t)$.
Here, 
\begin{equation}
\int\delta n_{j}({\bf r}, t) d{\bf r} = 0, \label{eq:delta_n}
\end{equation} 
because the total number of particles of both components are conserved.
Then,  $E_{{\rm int}, 1}(t)$ can be calculated as
\begin{equation}
\begin{array}{l}
 E_{{\rm int}, 1}(t) = \frac{g}{2}\int (n_{{\rm uni},1} + \delta n_{1}({\bf r}, t))^{2} d{\bf r} \\
  \\
 \ \ \ \ \ \ \ \ \ \ \ \ =  \frac{g}{2}\int n_{{\rm uni},1}^{2} d{\bf r} + \frac{g}{2}\int \delta n_{1}({\bf r}, t)^{2} d{\bf r} \\
   \\
 \ \ \ \ \ \ \ \ \ \ \ \ \geq \frac{g}{2}\int n_{{\rm uni},1}^{2} d{\bf r}. \label{eq:2C_Eint_change}
\end{array}
\end{equation} 
Thus, the emission and spread of the density waves increase $E_{{\rm int}, 1}(t)$.

In the two-component system, the energies $E_{j}$ of both components increase, although the pair of component 1 gets annihilated.
Although this seems unusual, this is sustained in the two-component system.
The annihilation of the pair in component 1 excites the density waves in component 2 because of the intercomponent interaction as well as those in component 1.
Here, the intercomponent interaction is repulsive; consequently, the local increase in the density of component 1 pushes component 2 from there and vice versa to yield $\delta n_{1}({\bf r}, t)\delta n_{2}({\bf r}, t) < 0$ almost everywhere (Fig. \ref{fig:2C_5.75_64_n_p}(c) and (i)).
Then, $E_{{\rm int}, 12}(t)$ decreases compared with the uniform system because
\begin{equation}
\begin{array}{l}
E_{{\rm int}, 12}(t) = g_{12}\int (n_{{\rm uni},1} + \delta n_{1}({\bf r}, t))(n_{{\rm uni},2} + \delta n_{2}({\bf r}, t)) d{\bf r} \\
  \\
 \ \ \ \ \ \ \ \ \ \ \ \ =  g_{12}\int n_{{\rm uni},1}n_{{\rm uni},2} d{\bf r} + g_{12}\int \delta n_{1}({\bf r}, t) \delta n_{2}({\bf r}, t) d{\bf r} \\
   \\
 \ \ \ \ \ \ \ \ \ \ \ \ \leq g_{12}\int n_{{\rm uni},1}n_{{\rm uni},2} d{\bf r}. \label{eq:2C_E12_change}
\end{array}
\end{equation} 
Through this process, the total energy of the system is conserved, and $E_{1}$ and $E_{2}$ increase and $E_{{\rm int}, 12}$ decreases until $t \simeq 300$ (Fig. \ref{fig:2C_5.75_64_d_E}(a)), and then they become statistically steady.


\subsection{Recurrence of a vortex pair}\label{sec:recursion}

In the small system with $g_{12} = 0.9g$, we find the creation of the rarefaction pulse and the recreation of the vortex pair.
Here, we call this recreation the recurrence of the vortex pair.
However, the recurrence does not occur in the medium system with $g_{12} = 0.9g$, although the rarefaction pulse appears.
These results are very different from those of the large system.


\begin{center}
\begin{figure}[h]
\includegraphics[width=18pc, height=15.75pc]{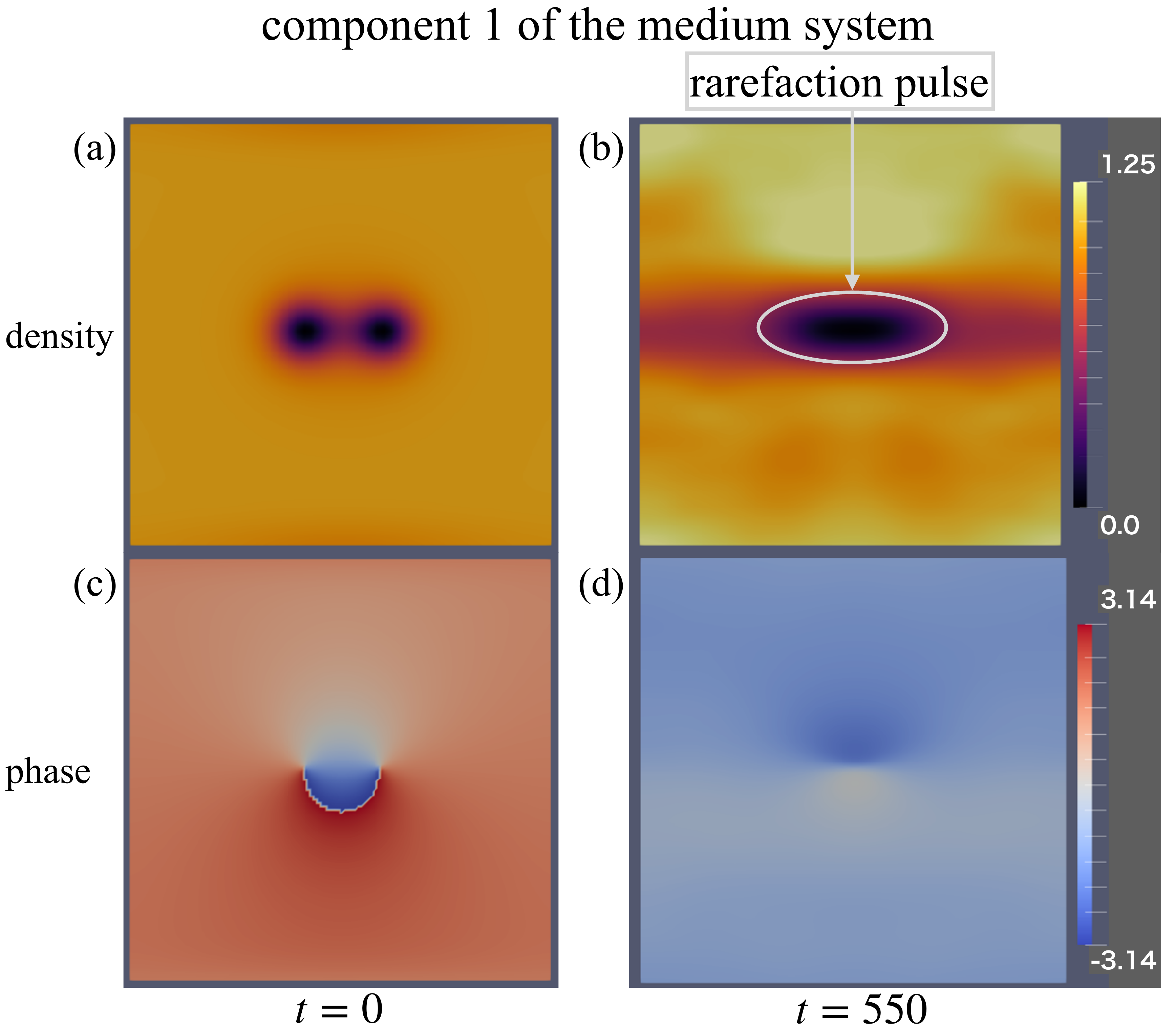}
\includegraphics[width=18pc, height=15.75pc]{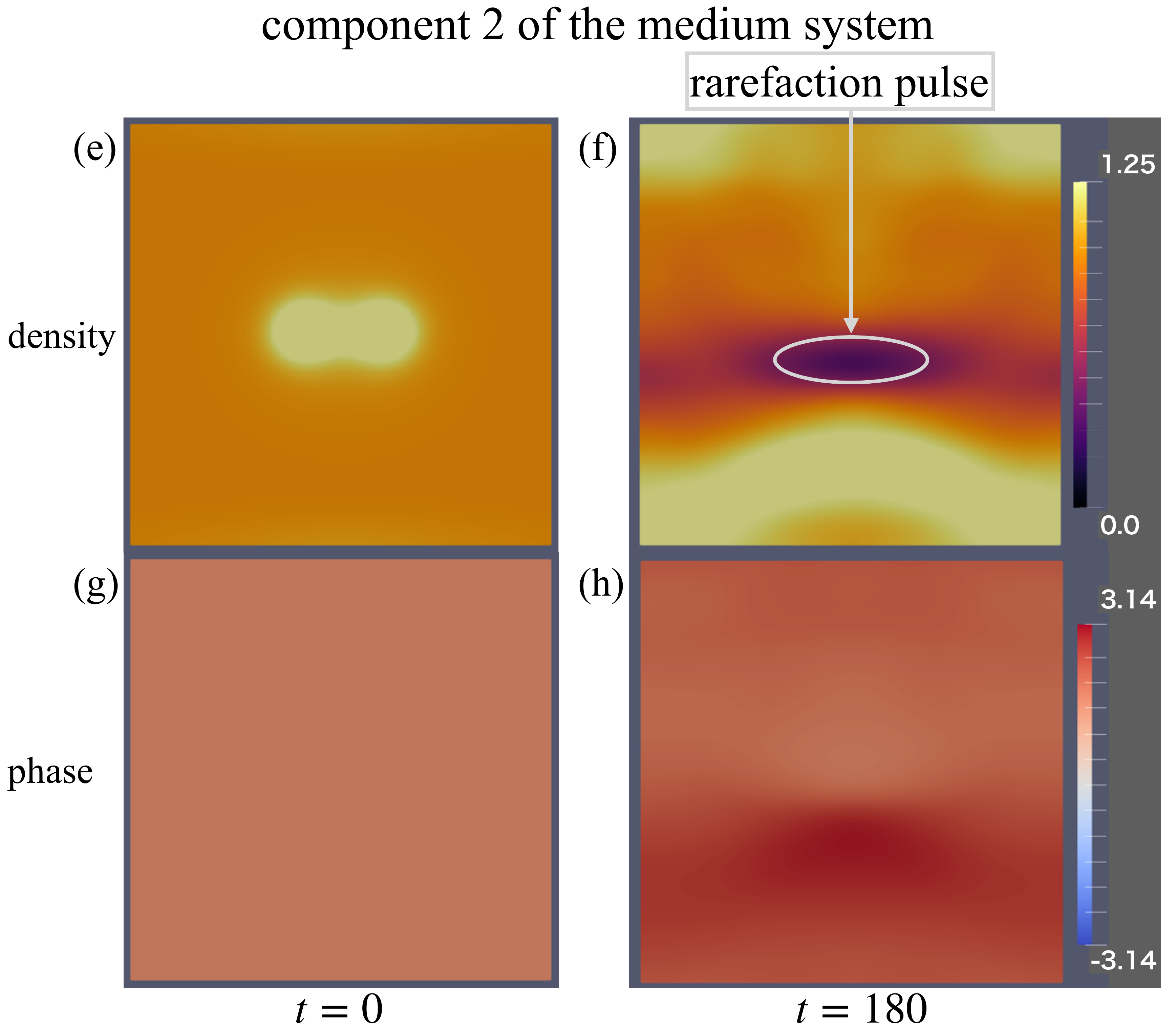}
\caption{\label{fig:2C_5.75_32_n_p} 
Dynamics when a rarefaction pulse is created in both components  in the medium system of $L = 32$ with $g_{12} = 0.9g$.
(a)--(d) describe component 1, and (e)--(h) describe component 2.
(a), (e), (c), and (g) are the density and phase at $t = 0$; (b) and (d) are those at $t = 550$; and (f) and (h) are those at $t = 180$.
The color bar of the density is normalized by the uniform density, and the range is $0$ -- $1.25$.
The white ellipses and arrows denote the rarefaction pulses.}
\end{figure}
\end{center}



\begin{center}
\begin{figure}[h]
\includegraphics[width=20pc, height=8.33pc]{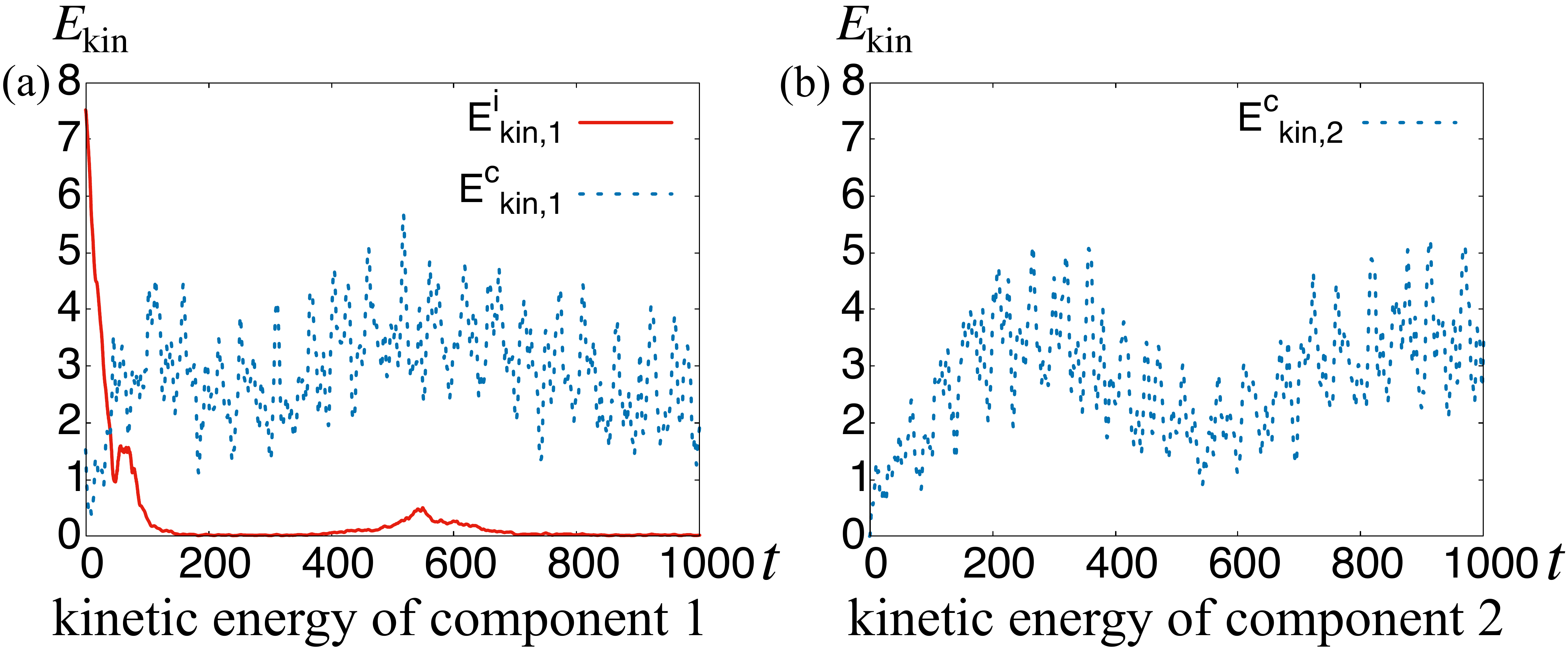}
\caption{\label{fig:2C_5.75_32_d_E} 
Time development of (a) $E_{{\rm kin}, 1}^{\rm i}(t)$ and $E_{{\rm kin}, 1}^{\rm c}(t)$, and (b) $E_{{\rm kin}, 2}^{\rm c}(t)$ in the medium system with $g_{12} = 0.9g$.
(a) shows the time development of the kinetic energies of component 1 and (b) shows that of component 2.
The denotations of each line are same as those in Fig. \ref{fig:2C_5.75_64_d_E}}
\end{figure}
\end{center}


Figures \ref{fig:2C_5.75_32_n_p} and \ref{fig:2C_5.75_32_d_E} show the density, phase, and time development of the energies in the medium system with $g_{12}=0.9g$.
The pair gets annihilated at $t \simeq 100$ like in the large system, and $E_{{\rm kin}, 1}^{\rm i}(t)$ decreases (Fig. \ref{fig:2C_5.75_32_d_E}(a)).
Then, the density waves spread over the system.
Thereafter, the density waves focus on a narrow region and gradually develop into a rarefaction pulse.
During the process, $E_{{\rm kin}, 1}^{\rm c}(t)$ and $E_{{\rm kin}, 1}^{\rm i}(t)$ increase until $t \simeq 550$, when the rarefaction pulse appears clearly with $n_{1}({\bf r}, t) \sim 10^{-2}$ and the local phase changes $\pi$, as shown in Fig. \ref{fig:2C_5.75_32_n_p}(b) and (d).
This rarefaction pulse subsequently decays, and finally, the density waves just spread.
In component 2, the rarefaction pulse also appears at $t \simeq 180$, but $E_{{\rm kin}, 2}^{\rm i}(t)$ shows a negligible increase (Fig. \ref{fig:2C_5.75_32_d_E}(b)) because the rarefaction pulse is shallow ($n_{2}({\bf r}, t) \sim 10^{-1}$) and the local change in the phase is less than $\pi$ (Fig. \ref{fig:2C_5.75_32_n_p}(f) and (h)).
Here, Fig. \ref{fig:2C_5.75_32_d_E}(b) only shows $E_{{\rm kin}, 2}^{\rm c}(t)$ because $E_{{\rm kin}, 2}^{\rm i}(t) \simeq 0$.
This shallow rarefaction pulse decays.
After these processes, the system eventually enters the statistical steady state where the density waves just spread in both components.


\begin{center}
\begin{figure}[h]
\includegraphics[width=18pc, height=15.75pc]{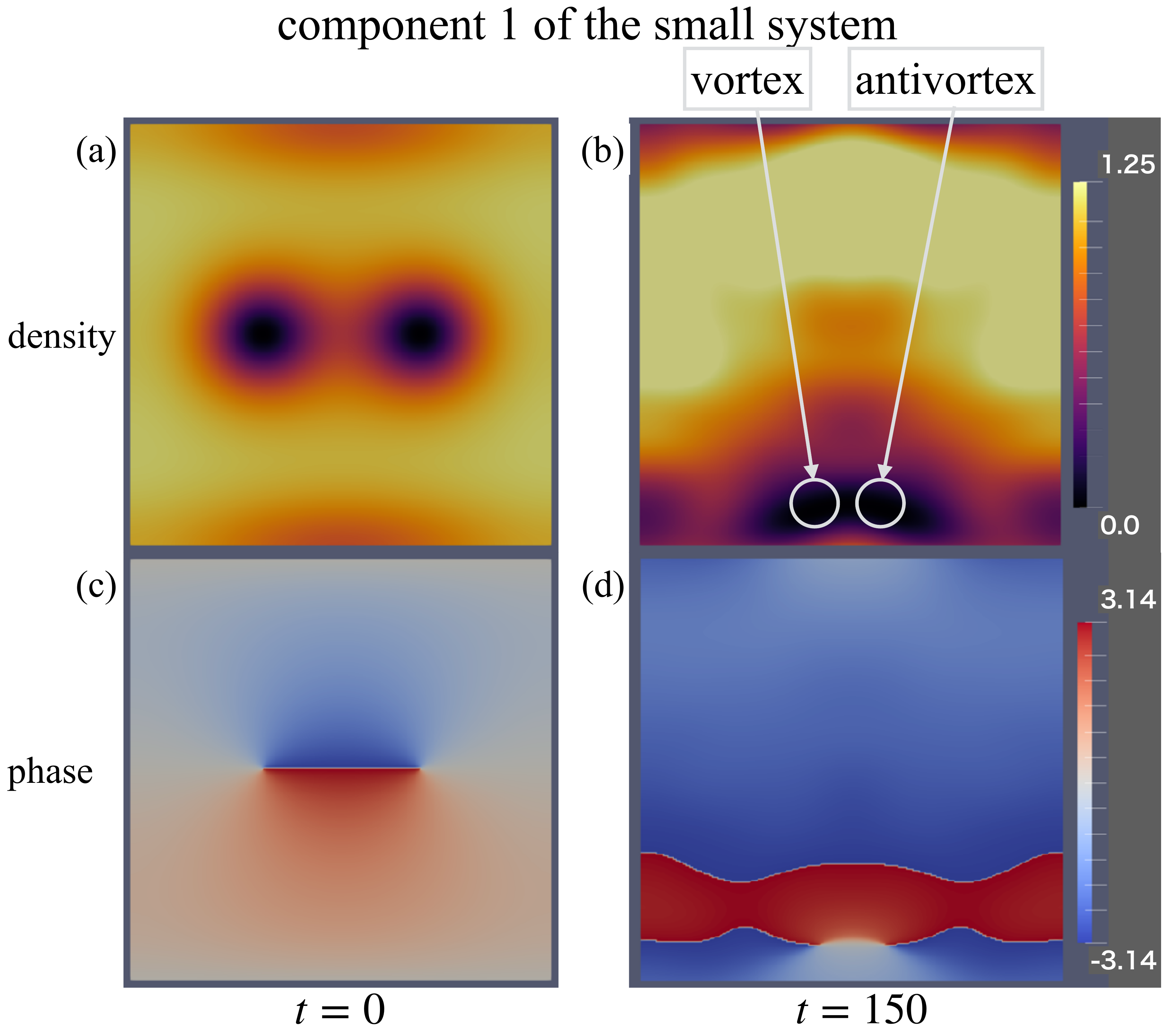}
\includegraphics[width=18pc, height=15.75pc]{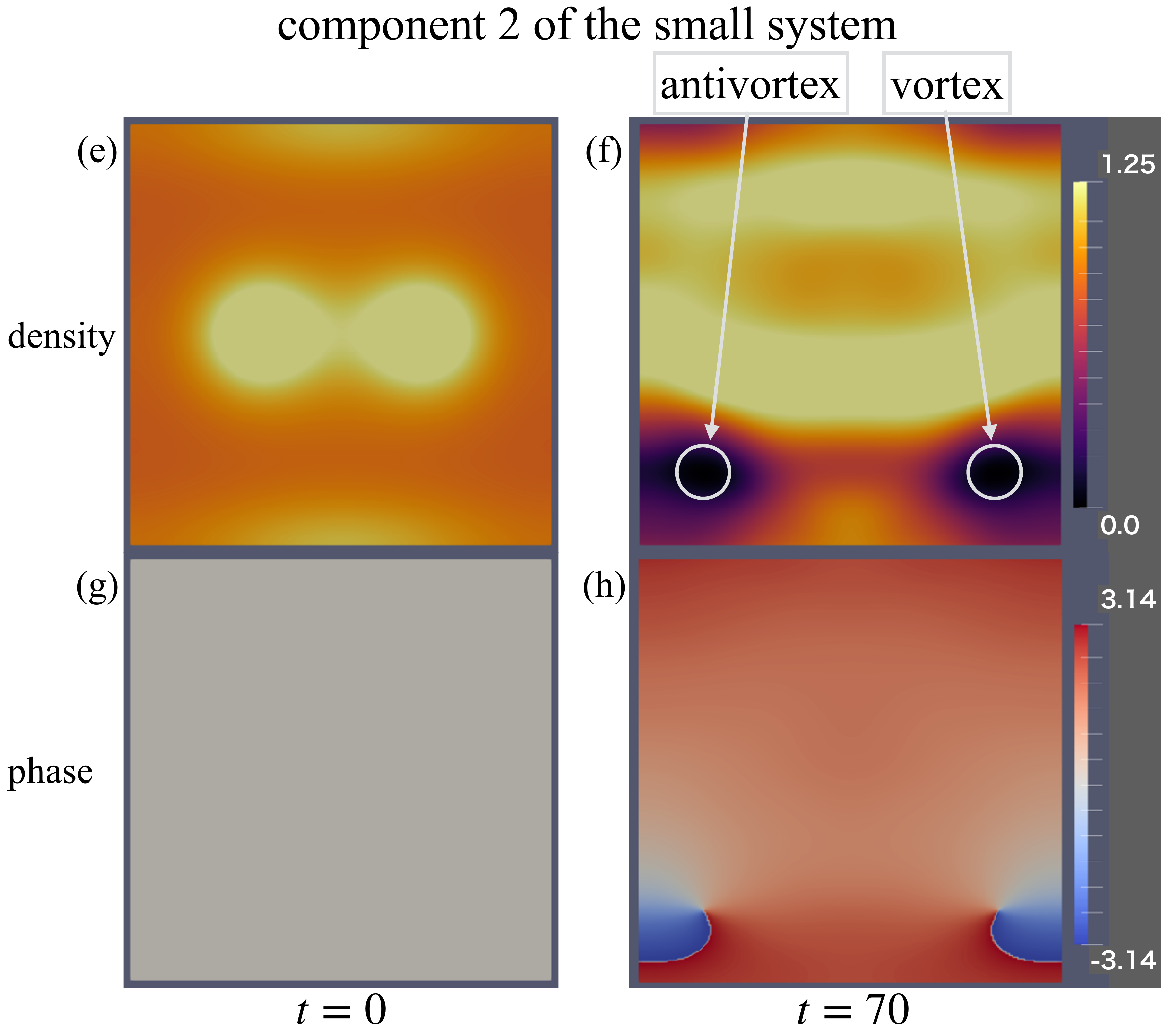}
\caption{\label{fig:2C_5.75_16_n_p} 
Dynamics when a vortex pair is recreated in both components in the small system of $L = 16$ with $g_{12} = 0.9g$.
(a)--(d) describe component 1, and (e)--(h) describe component 2.
(a), (e), (c), and (g) are the density and phase at $t = 0$; (b) and (d) are those at $t = 150$; and (f) and (h) are those at $t = 70$.
The color bar of the density is normalized by the uniform density, and the range is $0$ -- $1.25$.
The white circles and arrows denote the vortex and antivortex, respectively.}
\end{figure}
\end{center}



\begin{center}
\begin{figure}[h]
\includegraphics[width=20pc, height=15pc]{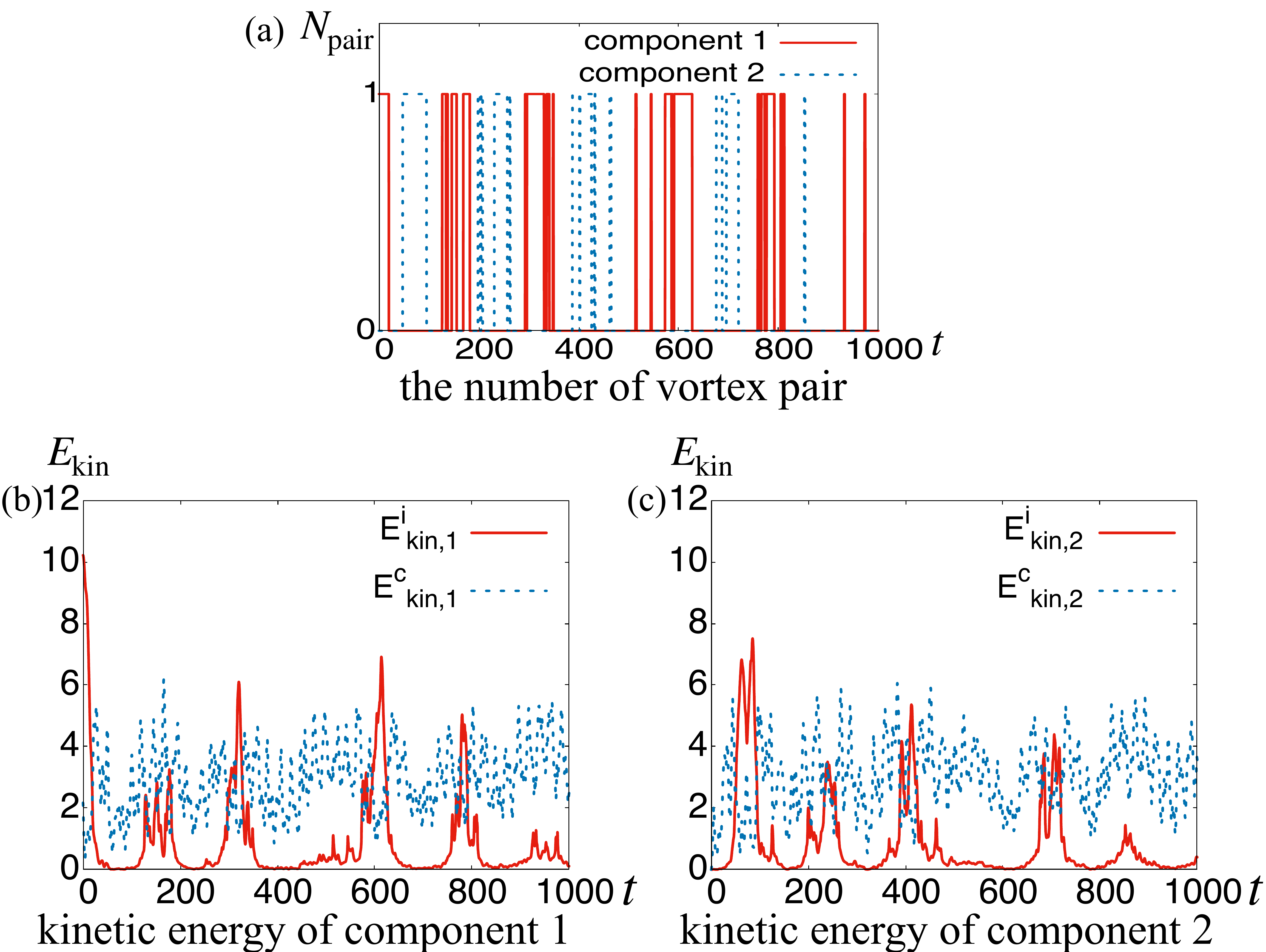}
\caption{\label{fig:2C_5.75_16_d_E} 
Time development of (a) the number $N_{\rm pair}(t)$ of vortex pairs, (b) $E_{{\rm kin}, 1}^{\rm i}(t)$ and $E_{{\rm kin}, 1}^{\rm c}(t)$, and (c) $E_{{\rm kin}, 2}^{\rm i}(t)$ and $E_{{\rm kin}, 2}^{\rm c}(t)$ in the small system with $g_{12} = 0.9g$.
The denotations of each line are the same as those in Fig. \ref{fig:2C_5.75_64_d_E}}
\end{figure}
\end{center}


Figure \ref{fig:2C_5.75_16_n_p} and \ref{fig:2C_5.75_16_d_E} present the results of the small system with $g_{12}=0.9g$.
The rarefaction pulse is created in component 2 after the initial annihilation of a pair in component 1, and this breaks up into two vortices immediately (Fig. \ref{fig:2C_5.75_16_n_p}(f) and (h), $t = 70$).
This pair also gets annihilated soon. 
After the annihilation in component 2, another pair is recreated in component 1 (Fig. \ref{fig:2C_5.75_16_n_p}(b) and (d), $t = 150$).
In this way, two components keep creating a vortex pair alternately.
Figure \ref{fig:2C_5.75_16_d_E}(a) shows the time development of the number of vortex pairs in each component, where a vortex is defined from the phase profile as a topological defect with quantized circulation.
When there is a vortex pair in component $j$, $N_{{\rm pair}, j} = 1$, and this state appears alternately between two components with a relatively long period (about $150$).
In each period, $N_{{\rm pair}, j}(t)$ can change from unity to $0$ a few times because a vortex pair can change continuously to a rarefaction pulse with no topological singularity.
The change from a pair to a rarefaction pulse yields $N_{{\rm pair}, j} = 0$.
The reverse change can occur, and thus, $N_{{\rm pair}, j}(t)$ alternates between unity and $0$ in a period.
We call the alternate creation of a vortex pair between two components the recurrence.
The recurrence is repeated several times (Fig. \ref{fig:2C_5.75_16_d_E}(a)).

\begin{center}
\begin{figure}[h]
\includegraphics[width=18pc, height=14.33pc]{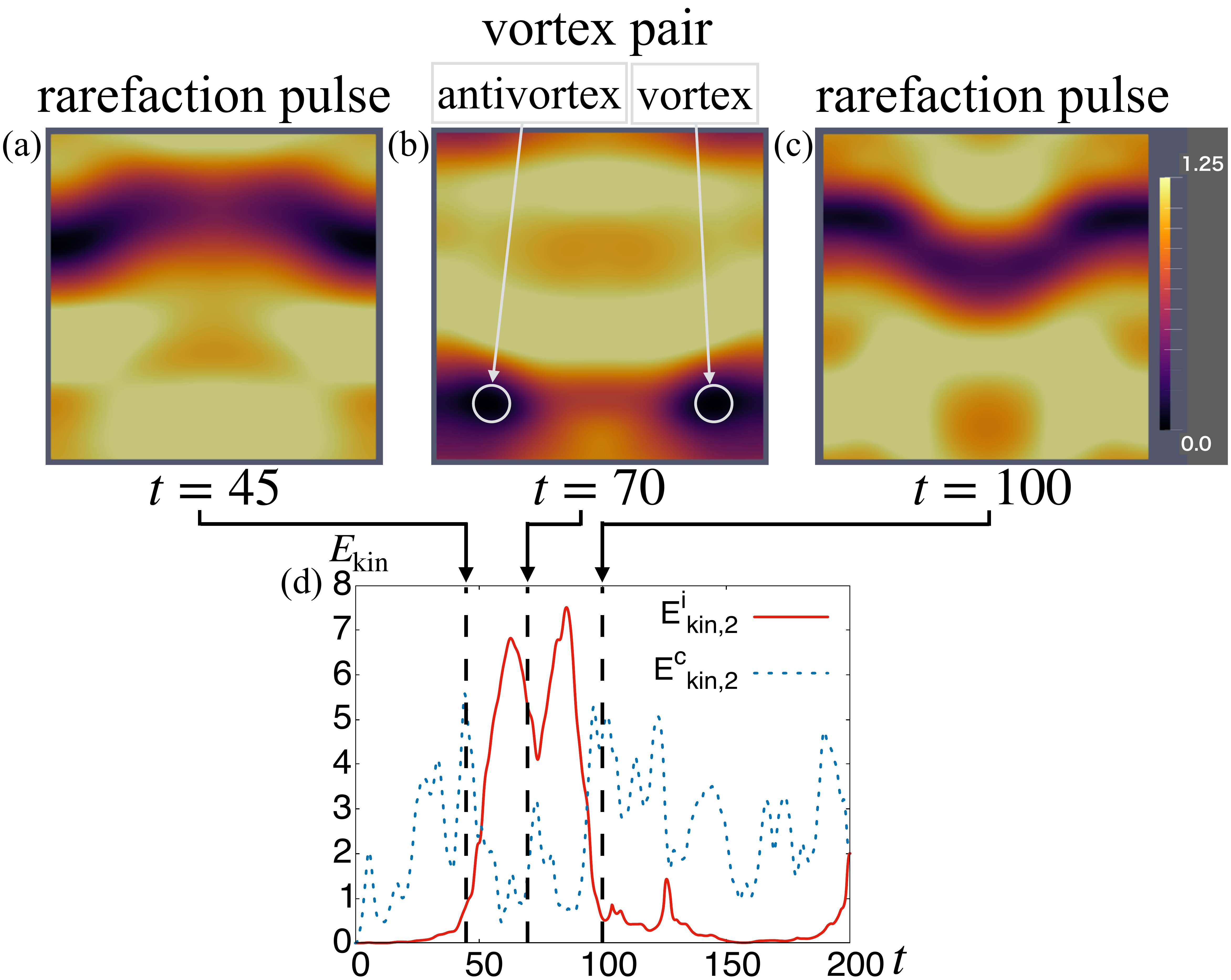}
\caption{\label{fig:2C_period}
Dynamics and development of the kinetic energies within a period in component 2 in the small system of $L = 16$ with $g_{12} = 0.9g$.
(a)--(c) show the densities at $t = 45$, $t = 70$, and $t = 100$, respectively.
(d) shows the time development of $E_{{\rm kin},2}^{\rm i}(t)$ and $E_{{\rm kin},2}^{\rm c}(t)$ from $t = 0$ to $t = 200$.
The vertical dashed black lines correspond to the time frames shown in (a)--(c).}
\end{figure}
\end{center}

The recurrence is also found in the characteristic time development of energy components.
Figure \ref{fig:2C_period} shows the dynamics and time development of the kinetic energies in component 2 from $t = 0$ to $t = 200$, and the first period of the recurrence is included in the duration.
Initially, $E_{{\rm kin},2}^{\rm c}(t)$ increases and a rarefaction pulse appears at $t \lesssim 45$ (Fig. \ref{fig:2C_period}(a) and (d)).
Then, $E_{{\rm kin},2}^{\rm c}(t)$ starts to change to $E_{{\rm kin},2}^{\rm i}(t)$ and a vortex pair is created around $t = 70$ (Fig. \ref{fig:2C_period}(b) and (d)).
Finally, the vortex pair returns to a rarefaction pulse and $E_{{\rm kin},2}^{\rm i}(t)$ changes to $E_{{\rm kin},2}^{\rm c}(t)$ around $t = 100$ (Fig. \ref{fig:2C_period}(c) and (d)).
Thereafter, both $E_{{\rm kin},2}^{\rm c}(t)$ and $E_{{\rm kin},2}^{\rm i}(t)$ become small around $t = 150$, and a vortex pair appears in component 1 (Fig. \ref{fig:2C_period}(d) and \ref{fig:2C_5.75_16_n_p}(b) and (d)).
These processes are repeated alternately within each period in each component, and finally, the recurrence slowly fades out and vanishes after $t \simeq 1000$.


\subsection{Distribution of the compressible and the incompressible kinetic energy}\label{sec:distribution_kinetic_energy}

\begin{center}
\begin{figure}[h]
\includegraphics[width=18pc, height=16.435pc]{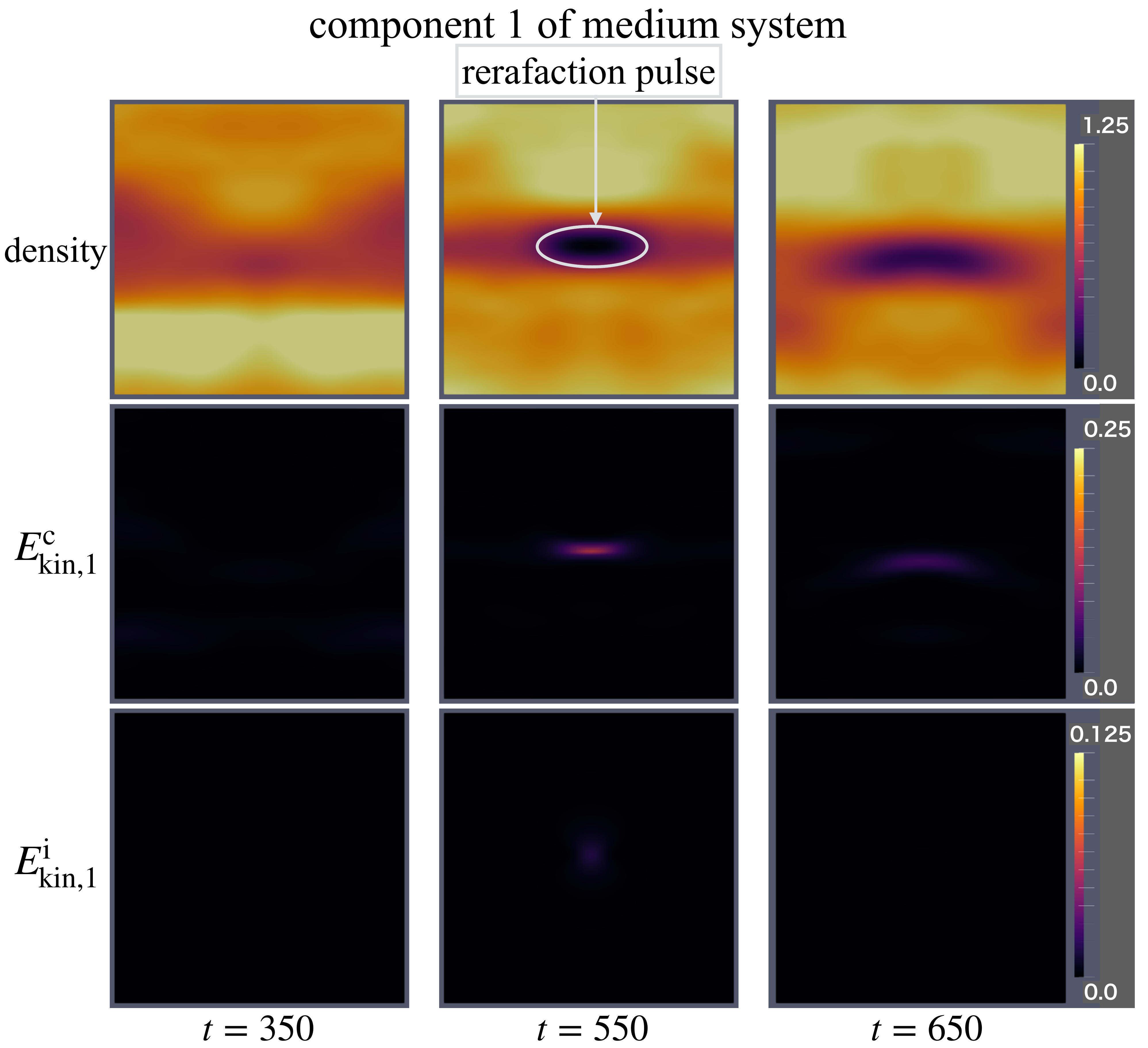}
\caption{\label{fig:2C_C1_32_kinetic_distribution}
Distributions of the density (top row), compressible kinetic energy $E_{{\rm kin}, j}^{\rm c}$ (middle row), and incompressible kinetic energy $E_{{\rm kin}, j}^{\rm i}$ (bottom row) of component 1 of the medium system.
The left, center, and right columns show the results at $t = 350$, $t = 550$, and $t = 650$, respectively.
The ranges of the color bars of each quantity are $0$ -- $1.25$ (density), $0$ -- $0.25$ ($E_{{\rm kin}, j}^{\rm c}$), and $0$ -- $0.125$ ($E_{{\rm kin}, j}^{\rm i}$).}
\end{figure}
\end{center}

\begin{center}
\begin{figure}[h]
\includegraphics[width=18pc, height=16.435pc]{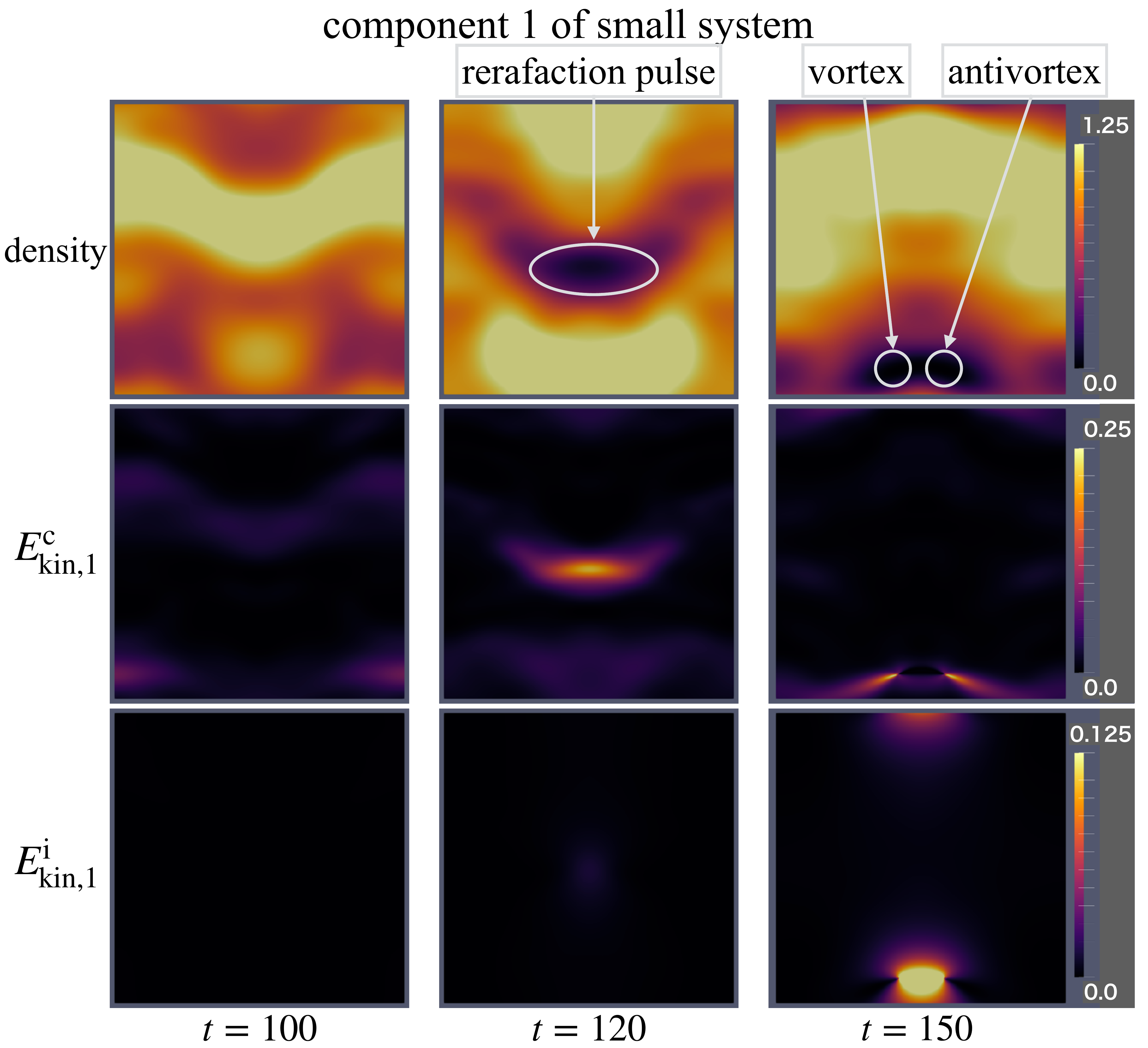}
\caption{\label{fig:2C_C1_16_kinetic_distribution}
Distributions of the density (top row), compressible kinetic energy $E_{{\rm kin}, j}^{\rm c}$ (middle row), and incompressible kinetic energy $E_{{\rm kin}, j}^{\rm i}$ (bottom row) of component 1 of the small system.
The row data and the ranges of the color bars are the same as those in Fig. \ref{fig:2C_C1_32_kinetic_distribution}.
The left, center, and right columns show the results at $t = 100$, $t = 120$, and $t = 150$, respectively.}
\end{figure}
\end{center}

In these systems, only component 1 has a vortex pair at the beginning.
However, the exchange of energies by the intercomponent interaction causes 
the annihilation of the vortex pair and the spread of the density waves.
Thereafter, in the medium system, the rarefaction pulse is created and both $E_{{\rm kin},1}^{\rm c}(t)$ and $E_{{\rm kin},1}^{\rm i}(t)$ increase temporarily at about $t \simeq 550$, but decrease with the decay of the rarefaction pulse (Fig. \ref{fig:2C_5.75_32_d_E}(a)).
In the small system, the recurrence occurs; then, $E_{{\rm kin},j}^{\rm c}(t)$ increases and changes to $E_{{\rm kin},j}^{\rm i}(t)$ (Fig. \ref{fig:2C_5.75_16_d_E}(b) and (c)).
Here, we show the spatial distribution of $E_{{\rm kin},j}^{\rm c}(t)$ and $E_{{\rm kin},j}^{\rm i}(t)$ in Fig. \ref{fig:2C_C1_32_kinetic_distribution} and \ref{fig:2C_C1_16_kinetic_distribution} to confirm these processes.


When the density waves spread out, the compressible kinetic energy is widely distributed across the entire system.
In the medium system, $E_{{\rm kin}, 1}^{\rm c}$ is widely distributed (Fig. \ref{fig:2C_C1_32_kinetic_distribution}, $t = 350$).
Then, the values of $E_{{\rm kin}, 1}^{\rm c}$ are relatively small everywhere.
Over time, $E_{{\rm kin}, 1}^{\rm c}$ returns to focus on a narrow region, creating the rarefaction pulse ($t=550$).
However, the change from $E_{{\rm kin}, 1}^{\rm c}$ to $E_{{\rm kin}, 1}^{\rm i}$ is not very effective, and thus, the rarefaction pulse does not develop into a vortex pair and just decays ($t = 650$).
In the small system, the values of $E_{{\rm kin}, j}^{\rm c}$ are relatively large everywhere (Fig. \ref{fig:2C_C1_16_kinetic_distribution}, $t=100$).
Thus, $E_{{\rm kin}, j}^{\rm c}$ can concentrate locally more than in the medium system ($t = 120$), and the change from $E_{{\rm kin}, j}^{\rm c}$ to $E_{{\rm kin}, j}^{\rm i}$ is sufficient to recreate a vortex pair ($t = 150$).
This means that the smaller system causes a more active exchange between the compressible and the incompressible kinetic energy.
After the recreation of a pair, $E_{{\rm kin}, j}^{\rm c}$ concentrates at the vortex cores and $E_{{\rm kin}, j}^{\rm i}$ is distributed near the pair.

If the exchange of energy between $E_{{\rm kin}, 1}^{\rm i}$ and $E_{{\rm kin}, 2}^{\rm i}$ is perfect, the recurrence should continue forever with some constant period.
In actual processes of recurrence, however, the exchange is not perfect.
Initially, $E_{{\rm kin}, 1}^{\rm i}$ is dominant in the kinetic energy of the system and $E_{{\rm kin}, j}^{\rm c}$ coming from stopping the imaginary time evolution is negligible.
Through the annihilation of a vortex pair, $E_{{\rm kin}, 1}^{\rm i}$ changes to $E_{{\rm kin}, 1}^{\rm c}$, and the compressible kinetic energy is exchanged between $E_{{\rm kin}, 1}^{\rm c}$ and $E_{{\rm kin}, 2}^{\rm c}$ by the intercomponent interaction.
Finally, $E_{{\rm kin}, 2}^{\rm c}$ focuses on a narrow region and changes to $E_{{\rm kin}, 2}^{\rm i}$ with creating a vortex pair in component 2, while some $E_{{\rm kin}, j}^{\rm c}$ remains in both components.
This imperfection of exchange between $E_{{\rm kin}, 1}^{\rm i}$ and $E_{{\rm kin}, 2}^{\rm i}$ occurs at every step of recurrence, and this leads to the lack of exact periodicity in Fig. \ref{fig:2C_5.75_16_d_E}(a) and the eventual decay of recurrence.

When a vortex is created, the velocity of the super fluid exceeds the critical velocity.
The critical velocity is of the order of the sound velocity, which comes from the dispersion relation, in one-component BECs \cite{Sasaki10}.
The dispersion relation is also calculated in two-component BECs \cite{PethickSmith08}, and the critical velocity of two-component BECs is found to be of the order of the sound velocity.
In this system, the superfluid velocity becomes large in the narrow region $E_{{\rm kin}, j}^{\rm c}$ concentrates.
Thus, the maximum velocity of each component exceeds the sound velocity in the medium or small system when a rarefaction pulse or a vortex pair is created, but not in the large system.
These results are consistent with the estimation of the order of the critical velocity in two-component BECs.


\section{Conclusions} \label{sec:Conclusions}

We find that the annihilation and the recurrence of vortex pairs occur in two-component BECs without dissipation.
Here, each phenomenon is sensitive to the intercomponent-interaction parameter and the system size.
If the intercomponent interaction is weak, energy and annihilation exchange between two components is negligible.
However, a strong-enough intercomponent interaction changes the situation drastically and annihilates a pair.
The medium system induces the concentration of the compressible kinetic energy emitted by the annihilation, and then the rarefaction pulse is created.
However, the change from compressible to incompressible kinetic energy is not very effective, and thus, a vortex pair is not created and the rarefaction pulse just decays.
The small system also induces the concentration of the compressible kinetic energy emitted by the annihilation and changes it to the incompressible kinetic energy.
Then, the recurrence occurs several times.

In summary, for the recurrence to occur, the annihilation of the vortex pair without dissipation and the concentration of the compressible kinetic energy are required.
The latter is certainly caused by the periodic boundary condition and the small system size.
However, the recurrence never occurs without the strong repulsive intercomponent interaction, which causes the annihilation of vortex pairs without dissipation and the exchange of energies between two components.
Hence, the recurrence is characteristic of two-component BECs.

Here, the annihilation of vortex pair, which is required for the recurrence, also depends on the initial distance between the vortices, and the distance can be changed depending on when imaginary time evolution is stopped.
In this study, we confine the distance to $5.75$ and confirm that the annihilation dose not occur when $g_{12}  \lesssim 0.8g$.
It is also confirmed that the critical distance below which the annihilation occurs becomes small at the limit of $g_{12} \rightarrow 0$, because the system becomes two independent condensates at that limit \cite{one_component}.
Thus, the critical distance is the increasing function of $g_{12}$.
The more detailed dependence of the annihilation on the distance is a future work.

We should comment on the time scale of a period of the recurrence in Fig. \ref{fig:2C_period}.
In component 2, $E_{{\rm kin}, 2}^{\rm i}$ increases and a rarefaction pulse appears about $t \lesssim 45$, and a vortex pair returns to another rarefaction pulse around $t=100$; the period of the recurrence is about $60$ in the early time.
The time scale of this period is comparable to the lifetime of the vortex pair, because $E_{{\rm kin},1}^{\rm i}(t)$ immediately increases after the vortex pair returns to a rarefaction pulse in component 2 around $t = 100$ (Fig. \ref{fig:2C_5.75_16_d_E}(b) and (c)).
The period just increases later around $t=500$ because of the imperfection of the exchange between $E_{{\rm kin}, 1}^{\rm i}$ and $E_{{\rm kin}, 2}^{\rm i}$, however, the order of the period is still similar.
The qualitative discussion of what parametrically sets this timescale would be important.


\begin{acknowledgments}
This work was supported by a Grant-in-Aid for JSPS Fellows (grant nos. 20J14459 and JP20H01855).
\end{acknowledgments}

\bibliography{book,paper}

\begin{thebibliography}{47}
\bibitem [{one()}]{one_component}%
  \BibitemOpen
  \href@noop {} {}\bibinfo {note} {In order to confirm the effect of the
  effective dissipation due to phonons, we calculate dynamics of a
  vortex-antivortex pair in one-component system without dissipation by GP
  equation. Here, we take the initial distance between two vortices is
  $1.75\xi$ when two vortex cores almost overlap. Then, the pair never
  annihilates for such a long time considered in this paper, thus the effective
  dissipation due to phonons is negligible in this study}\BibitemShut {NoStop}%
\end{thebibliography}

\end{document}